\documentclass{ws-ijmpd}
\usepackage{graphicx,amsmath,amssymb}
\usepackage{color}

\def\openone{\leavevmode\hbox{\small1\kern-3.1pt\normalsize1}}
\begin{document}

\markboth{D. V. Ahluwalia-Khalilova}{Dark matter, and its darkness}

\title{DARK MATTER, AND ITS DARKNESS}

\author{D. V. AHLUWALIA-KHALILOVA}

\address{Department of Physics and Astronomy, Rutherford Building\\ 
University of Canterbury, Private Bag 4800\\ 
Christchurch 8020, New Zealand\\
E-mail: dharamvir.ahluwalia-khalilova@canterbury.ac.nz}

\maketitle

\begin{abstract} 
Assuming the validity of the general relativistic description of
gravitation on astrophysical and cosmological length scales, we
analytically infer that the Friedmann-Robertson-Walker cosmology with
Einsteinian cosmological constant, and a vanishing spatial curvature
constant, unambiguously requires significant amount of dark
matter. This requirement is consistent with other indications 
for dark matter. The same
spacetime symmetries that underlie the freely falling frames of
Einsteinian gravity also provide symmetries, which for
the spin one half representation space, furnish a novel construct that
carries extremely limited interactions with respect to the terrestrial
detectors made of the standard model material. Both the `luminous' and
`dark' matter turn out to be residents of the same representation
space but they derive their respective `luminosity' and `darkness'
from either belonging to the sector with $(CPT)^2 = + \openone$,
or to the sector with $(CPT)^2 = - \openone$.
\end{abstract}


\section{Introduction}
The first task that we set for ourselves in this essay 
is to show that the Friedmann-Robertson-Walker (FRW) cosmology with 
$k=0$ and $\omega^\Lambda = -1$ requires significant amount of dark matter.
This we accomplish without 
invoking the non-Keplerian galactic rotational curves or gravitational
lensing data~[\refcite{Persic:1995ru}--\refcite{Hoekstra:2003pn}].
The result can be interpreted in two ways. As a complimentary requirement
for the dark matter, or as a breakdown of 
the general relativistic description of gravitation 
at large astrophysical and cosmological 
scales~[\refcite{Milgrom:1983pn}--\refcite{Sobouti:2006rd}].
Continuing the study with the former possibility,  we provide a natural 
explanation for the darkness of dark matter. This we  do by showing
that the  
Dirac framework may suffice only for the description of the standard model
fermions, and that dark matter \textemdash~if we consider it fermionic, with
spin one half  \textemdash~naturally asks for a dramatically new description
where the dark matter field carries a Klein-Gordon, and not Dirac, propagator.
This thesis is a natural consequence of a recent preprint
[\refcite{Ahluwalia-Khalilova:2006ez}] and two
publications late last 
year~[\refcite{Ahluwalia-Khalilova:2004sz,Ahluwalia-Khalilova:2004ab}]. 
The latter publications 
are somewhat technical, here we briefly reformulate them 
in a physically accessible manner.
Should these results appear surprising
then it suffices to note that while the abstract
of Ref.~[\refcite{Ahluwalia-Khalilova:2004sz}] began 
as, ``We report an unexpected theoretical discovery of a 
spin one-half matter field with mass dimension one,'' 
the abstract of Ref.~[\refcite{Ahluwalia-Khalilova:2004ab}]
opened with the sentence, ``We provide the 
first details of the unexpected theoretical 
discovery of a spin-one half matter field with mass 
dimension one.''

For making this essay easily accessible to younger students
we adopt a pedagogical style. A more knowledgeable reader may
in some instances just smile, and proceed without further 
interruption.\footnote{To strike a balance between these two competing 
needs Sec.~\ref{Sec:2} is significantly more detailed than
Sec.~\ref{Sec:3}. This has been done because the main thesis of 
  Sec.~\ref{Sec:3} has recently appeared in 
Ref.~[\refcite{Ahluwalia-Khalilova:2004ab}] in  monographic detail.}

\section{Dark matter: A phenomenological existence proof \label{Sec:2}} 
Since 1997, data on supernovae 1a luminosities has built an impressive
evidence for the existence of a form of energy that is accelerating
the present expansion of the 
universe~[\refcite{Perlmutter:1997zf}--\refcite{Astier:2005qq}]. 
The latest 
data analysis~[\refcite{Astier:2005qq}] 
favors this dark energy to be 
of the form postulated by Einstein in his now well-known 
cosmological term. The previous conflict between SN1a data and the 
WMAP data on anisotropies in the cosmic microwave background noted by 
Jassal, Bagla, and Padmanabhan~[\refcite{Jassal:2005qc}]
is now resolved~[\refcite{Jassal:2006gf}].
Furthermore, observations on primordial cosmic microwave background and
large scale structure also favor a dark energy dominated universe 
which is spatially flat~[\refcite{Spergel:2003cb,Copeland:2006wr}].

In the context of inhomogeneous cosmologies, 
Wiltshire has raised a serious objection~[\refcite{Wiltshire:2005fw}, also
\refcite{Alnes:2006uk}--\refcite{Rasanen:2006kp}] 
to the just-stated standard interpretation of
the SN 1a data and has provided an alternate 
interpretation which does not require dark energy~[\refcite{f2}].
Working within the inflationary paradigm, 
he carefully distinguishes various clocks rates such as those
associated with the astrophysically large gravitationally bound systems and
the cosmic voids. 
Conceptually, the raised questions can hardly be ignored. Yet,
the final verdict shall undoubtedly depend
on whether such considerations turn out to be quantitatively
important. Preliminary calculations  
are already encouraging~[\refcite{Wiltshire:2005fw,Carter:2005mu}]. 
So, if here we proceed with dark energy as an option
it is in the spirit of an open enquiry without prejudice as to
the ultimate nature of the physical reality.

In the canonical wisdom,
the data thus favors a FRW cosmology where $k=0,\omega^\Lambda= -1$ with
$\rho^\Lambda$ as constant. However,
with an exception of a few 
papers, 
it is rarely appreciated that in this cosmology the \textit{non-linearity}
of the Einstein field equations 
completely determines the proportionality
constant that appears in the  
scale factor $a(t) \propto \sinh^{2/3}\left([3\Lambda]^{1/2}\, t/ 2\right)$
~[\refcite{Ahluwalia-Khalilova:2006ez}--\refcite{Lake:2006fb}]. The
proportionality constant depends on densities associated with
matter and the Einsteinian cosmological constant $\Lambda$. The result 
of the analysis presented 
in~[\refcite{Ahluwalia-Khalilova:2006ez,Gron:2002ox}]
is that the ratio of the corresponding densities is not arbitrary but it carries a
well-determined temporal dependence. This leads to significant
strengthening of the standard paradigm of the $\Lambda\mathrm{CDM}$  
cosmology, i.e the FRW cosmology defined by the 
set~[\refcite{f3}].

\[
\{k=0,\; w^\Lambda=-1,\; \rho=\rho_\mathrm{m},
\;p=p_\mathrm{m}=0,\;\;\rho^\Lambda={\mathrm{constant}}\}\label{eq:set}
\]
 In particular,
the stated circumstance when exploited
in the context of the stellar age bound determined by the ages of the 
globular clusters, and the fact that 
fractional density associated with the standard model content 
$\Omega_\mathrm{sm} \approx 0.05$, requires existence of dark matter. These arguments are
now refined and extended exploiting the recent work 
of~[\refcite{Ahluwalia-Khalilova:2006ez}, 
also \refcite{Vishwakarma:2006dj,Ahluwalia-Khalilova:2006ky}].

For the stated cosmology the Einstein field
equations, when their non-linearity is fully respected~[\refcite{Ahluwalia-Khalilova:2006ez}], 
yield for the matter-dominated epoch

\begin{eqnarray}
&&  \Omega_\Lambda(t) +  \Omega_\mathrm{m}(t) =1 \label{eq:a}\\
&& \Omega_\Lambda(t) -  \zeta(t)\, \Omega_\mathrm{m}(t) =0 \label{eq:b} 
\end{eqnarray} 
where
\begin{equation}
\zeta(t):=\sinh^2
\left(\frac{\sqrt{3 }}{2} \frac{t}{\tau_\Lambda} \right)\label{eq:c}
\end{equation}
Here we introduced $\tau_\Lambda:= \sqrt{1/\Lambda}$.
The rest of the notation is standard, and 
is defined in 
Ref.~[\refcite{Ahluwalia-Khalilova:2006ez}] with the special attention being
paid to the fact that all matter components in $\Omega_\mathrm{m}$ are
considered non-relativistic (i.e., $p_\mathrm{m}=0$). The latter  is the import of
considering the `matter dominated epoch'. Specifically, the fractional
densities $\Omega_\Lambda$ and $\Omega_\mathrm{m}$ are defined as

\begin{equation} 
      \Omega_\Lambda := \frac{8 \pi G \rho^\Lambda}{3 H^2}
=
\frac{\Lambda}{3 H^2}\,,\qquad
\Omega_{\mathrm m} :=\frac{8\pi G \rho_\mathrm{m}}{3 H^2} 
      \label{eq:t4}
\end{equation}
where we used $\rho^\Lambda := \Lambda/(8 \pi G)$. Here, $G$ is the 
Newtonian gravitational constant
and equals 
$6.7065 \times 10^{-39}\; \mathrm{GeV}^{-2}$. 
And, $H$ is the Hubble parameter. Its 
value for the present cosmic epoch shall be represented by $H_0$ below.

Equations (\ref{eq:a}) and (\ref{eq:b}) carry the following solutions

\begin{equation}
\Omega_\Lambda = \frac{\zeta(t)}{1+\zeta(t)},\quad
\Omega_\mathrm{m} = \frac{1}{1+\zeta(t)}
\end{equation}
Use of the definition of $\zeta(t)$ from Eq.~(\ref{eq:c}) immediately 
renders the above result into the form

\begin{equation}
\Omega_\Lambda = \tanh^2
\left(\frac{\sqrt{3 }}{2} \frac{t}{\tau_\Lambda} \right),\quad
\Omega_\mathrm{m} = 
\mathrm{sech}^2\left(\frac{\sqrt{3 }}{2} \frac{t}{\tau_\Lambda} \right)
\label{eq:omega}
\end{equation}
Our task for the phenomenological existence proof of dark matter
now reduces  to showing that $t/\tau_\Lambda$ for the present
cosmic epoch is such that the resulting $\Omega_\mathrm{m}$ far exceeds 
that which can be accounted for by the standard model contribution of
$\Omega_\mathrm{sm} \approx 0.05$ 

As such,
in order to obtain these fractional densities for the present cosmic epoch
we now need to constrain the cosmological constant $\Lambda$ (that
yields $\tau_\Lambda$), and to obtain
an expression for the age of the universe within the considered FRW 
cosmology. We shall first present a detailed calculation. This would
be complimented by a `back of the envelop' estimate to complete the
phenomenological existence proof of dark matter.
That would then allow us to return to the important question
of its darkness.

\begin{figure}
\includegraphics{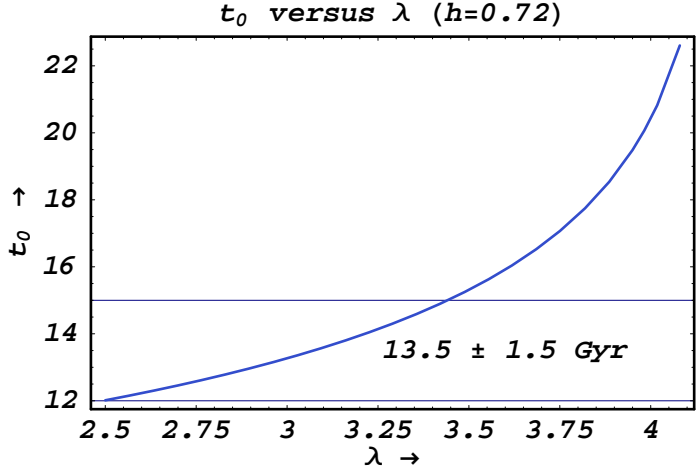}
\caption{
The present cosmic epoch of $13.5 \pm 1.5\;\mathrm{Gyr}$ corresponds
to $2.50 \le\lambda \le 3.45$ (for h=0.72). It can be shown that 
for h=0.64, the range reduces to roughly $1.3 \le \lambda\le  
2.4$, while for $h=0.80$ the present cosmic epoch of 
$13.5 \pm 1.5\;\mathrm{Gyr}$ corresponds to roughly 
$3.7 \le \lambda \le 4.6$. \label{Fig: t0X}}
\end{figure}

\begin{figure}
\noindent
\includegraphics{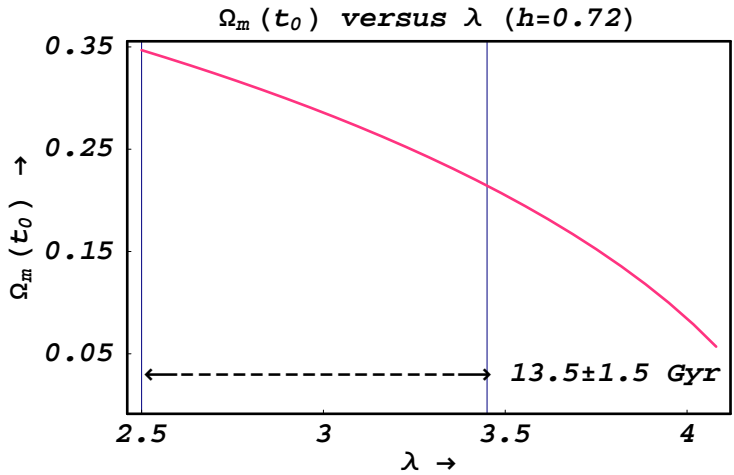}
\caption{The variation of $\Omega_\mathrm{m}$ with $\lambda$.
For the present cosmic epoch $0.22 \le \Omega_\mathrm{m} \le
0.35$. \label{Fig: OmegaMatterToday}}
\end{figure}

The scale factor $a(t)$ evaluated in 
Ref.~[\refcite{Ahluwalia-Khalilova:2006ez,Frieman:1994gf}] 
yields
the present age of the universe to be~[\refcite{f4}]

\begin{equation}
t_0 = \frac{2}{3 H_0} \frac{1}{\sqrt{\Omega_\Lambda(t_0)}}\ln\left[
\frac{1+\sqrt{\Omega_\Lambda(t_0)}}{\sqrt{1-\Omega_\Lambda(t_0)}}\right]
\end{equation}
This is in agreement, e.g., with the results contained 
in Eq.~3.32 of Ref.~[\refcite{Kolb:1990Turner}]
and in Eq.~54 of Ref.~[\refcite{Copeland:2006wr}]. In the expression 
for $t_0$, and  $\Omega_\mathrm{m}$ evaluated at $t_0$, 
we now substitute $\Omega_\Lambda(t_0):= \Lambda/(3 H_0^2)$, and set 

\begin{equation}
H_0^{-1} = 9.776 h^{-1}\;\mathrm{Gyr},\quad 0.64 \le h \le  0.80
\end{equation}
as obtained from the Hubble Space Telescope Key project~[\refcite{Freedman:2000cf}]. Concurrently, we take the cosmological 
constant to be

\begin{equation}
\Lambda = \lambda \, 8 \pi G \, 10^{-47}\;\mathrm{GeV}^4 \label{eq:lambda}
\end{equation}
where $\lambda$ is of the order of unity 
(to be constrained below).
This exercise gives

\begin{subequations}
\begin{eqnarray}
&& t_0 = \frac{18.55}{\sqrt{\lambda}} \mathrm{ln}
\left[\frac{3+ 1.054 \sqrt{\frac{\lambda}{h^2}} }
{\sqrt{9 - 1.111\, \frac{\lambda}{h^2}}}\right]\;\mathrm{Gyr} 
\label{eq:t0Gyr}\\
&& \tau_\Lambda = \frac{16.06}{\sqrt{\lambda}}\;\mathrm{Gyr}\\
&& \Omega_\mathrm{m}(t_0) =  \mathrm{sech}^2
\left[\frac{1.732}{\sqrt{\lambda}}\mathrm{ln}\left(
\frac{3+ 1.054 \sqrt{\frac{\lambda}{h^2}} }
{\sqrt{9 - 1.111\, \frac{\lambda}{h^2}}}\right)\right]
\end{eqnarray}
\end{subequations}
For $\lambda \ge 2.5$ and  $h=0.72$ (a suitably extended range of 
$\lambda$ and $h$ shall be discussed soon, below), 
the variations of $t_0$ and $\Omega_m(t_0)$ as a function of $\lambda$ are
given in Figures~\ref{Fig: t0X} and \ref{Fig: OmegaMatterToday}. 
It is apparent from Fig.~\ref{Fig: t0X} that 
the demand that $t_0$ be greater than 
the stellar age bound~[\refcite{Chaboyer84:2001,Copeland:2006wr}] of
$12\;\mathrm{Gyr}$, constrains the cosmological constant to 
$\lambda \ge 2.5$ (for $h=0.72$).
The present cosmic epoch of $13.5 \pm 1.5\;\mathrm{Gyr}$ 
corresponds to $2.50 \le \lambda \le 3.45$ (for h=0.72).

Figure \ref{Fig: OmegaMatterToday} 
clearly shows that for the present cosmic epoch $0.22 \le 
\Omega_\mathrm{m} \le
0.35$. 
 The standard model contribution $\Omega_\mathrm{sm} \approx 0.05$ falls
too short to account for the obtained $\Omega_\mathrm{m}$. This deficit is
the dark matter. Its existence is here  arrived at without 
invoking the non-Keplerian galactic rotational curves or the data on
gravitational 
lensing~[\refcite{Persic:1995ru}--\refcite{Hoekstra:2003pn}].

\begin{figure}
\includegraphics{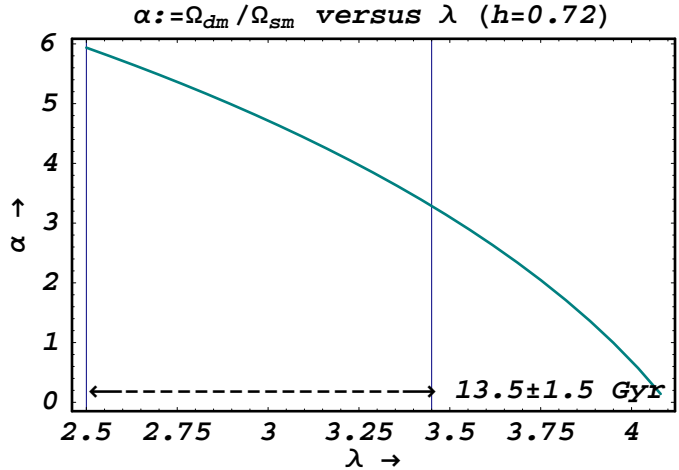}
\caption{The plot of $\alpha$ as a function 
of $\lambda$ for $\Omega_\mathrm{sm} = 0.05$; showing that
there is roughly three to six times more dark matter as 
compared to the total contribution from the particles of 
the standard model 
of high energy physics. \label{Fig: OmegaDarkMatterToday}}
\end{figure}

To emphasize this point further it is instructive to define

\begin{equation}
\alpha=\frac{\Omega_\mathrm{m} -\Omega_\mathrm{sm}}{\Omega_\mathrm{sm}} :=
\frac{\Omega_\mathrm{dm}}{\Omega_\mathrm{sm}}
\end{equation}
In Figure~\ref{Fig: OmegaDarkMatterToday},  
we exhibit  its variation with $\lambda$ for 
$\Omega_\mathrm{sm} = 0.05$. For the present cosmic epoch we thus
find $ 3.3 \le \alpha \le 5.9$. Or, equivalently
$0.17 \le \Omega_\mathrm{dm} \le 0.30$.
That is, for the present epoch, the cosmic energy budget carries
some three to six times as much dark matter as that contributed by
the particles of the standard model of the high energy physics.

\begin{figure}
\includegraphics{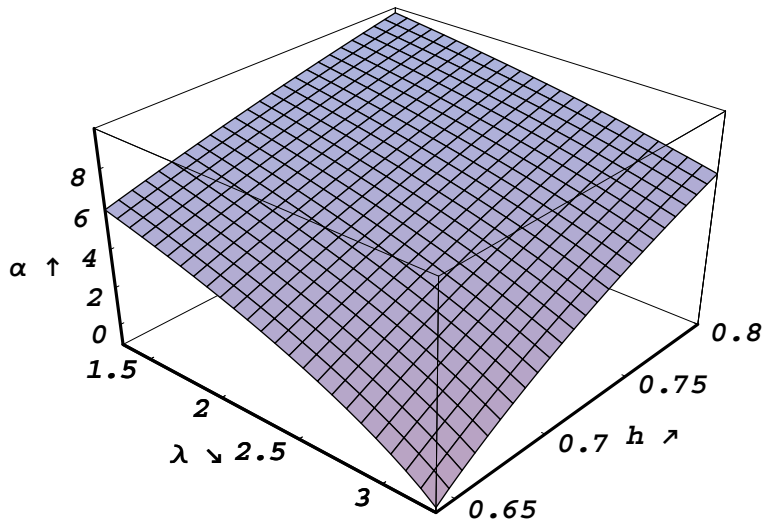}
\caption{The variation of $\alpha$ as a function 
of $\lambda$ and $h$ for $\Omega_\mathrm{sm} = 0.05$. It shows that
for almost the entire displayed range,  $ 1.3 \le \lambda \le
3.3$ and $0.64 \le h \le 0.80$, the dark matter measuring parameter
$\alpha$ remains significantly away from zero. Note parenthetically,
that the denominator of Eq.~(\ref{eq:t0Gyr}) yields physically meaningless
region just beyond $\lambda = 3.3$ around $h=0.64$. This is the region
at the lowest right hand corner of the above plot. And, only in this 
extreme
region does $\alpha$ hover around zero. \label{Fig: Alpha3D6480}}
\end{figure}

Much of the above considerations were confined to $h=0.72$. 
Figure~\ref{Fig: Alpha3D6480} 
shows that a similar result holds for the entire
observationally allowed range $0.64 \le h \le 0.80$. That is,
the arrived dark matter existence in the present cosmic epoch
is not specific to $h=0.72$. 

The whole analysis can be put in the following `back of the envelop'
argument. For $h=0.72$, the stellar age bound constrains the Einsteinian 
cosmological constant \textemdash~see Eq.~(\ref{eq:lambda})  \textemdash~ to
$\lambda \approx 3.0$. This gives $\tau_\Lambda \approx  
9.3 \;\mathrm{Gyr}$. For the present epoch of roughly 
$13.5\;\mathrm{Gyr}$ (corresponding to $\lambda \approx 3.0$), 
Eq.~(\ref{eq:omega}) yields $\Omega_\mathrm{m} \approx 0.28$.
However, $\Omega_\mathrm{sm} \approx 0.05$ can only account
for a small fraction of this obtained value. This leaves the fractional
density $\Omega_\mathrm{dm}:=\Omega_\mathrm{m} - \Omega_\mathrm{sm} \approx 0.23$
to point towards some form of \textit{non-standard} model matter 
in the non-relativistic form. This, by definition, is 
the astrophysical/cosmic dark matter~[\refcite{f5}].

The dramatic discordance between 
the calculated $\Omega_\mathrm{m}$ and the observed 
$\Omega_\mathrm{sm}$ may be alternatively interpreted,
not as pointing towards the existence of some
non-standard model matter (i.e., dark matter), but 
as a breakdown of the general relativistic description of gravitation 
at large astrophysical and cosmological length scales. Such a degeneracy in possible
interpretations is best resolved by experimental and observational efforts.
In the remainder of this essay we shall follow the dark-matter interpretation
and find that it may indeed be pointing towards a very novel
form of matter.

\section{Dark matter: Its darkness \label{Sec:3}}
All fermionic 
fields of the standard model of particle
physics may be expressed in terms of the Dirac spinors.
Formally, these spinors live in the $(1/2,0)\oplus(0,1/2)$
representation space~[\refcite{Weinberg:1995sw,Ryder:1996lh}].
Since the 1962 work of Wigner  it is known that these may not describe the 
entire physical reality of the high energy physics~
[\refcite{Wigner:1964ep}].
Yet, till late last year,
mostly due to non-trivial technical reasons and for an apparent lack 
of physical motivation (i.e, before
the suggestions of dark matter was taken seriously),  
no concrete construct of the type suggested by Wigner 
existed for spin one half~[\refcite{f6}].

To heuristically arrive at the conclusion of
Wigner's notes, as he called them, 
we first make the obvious observation that
the indicated representation space supports various symmetries.
Of these, the charge conjugation ($C$), the parity ($P)$, and
time reversal ($T$) symmetries are the most relevant for the argument. The 
Dirac spinors are the eigenspinors of the $P$ operator. Now 
$C$ and $P$ when acting upon these spinors do not commute; instead, 
they anti-commute. Consequently, a quantum field built upon the eigenspinors
of the $C$ operator should describe physically distinct 
type of  new particles.

A brief report on the construction of such a quantum field was recently 
presented in~[\refcite{Ahluwalia-Khalilova:2004sz}] while the 
monographic details were archived 
in~[\refcite{Ahluwalia-Khalilova:2004ab}].
The ensuing result states that the quantum field based on the
eigenspinors of the $C$ operator does not carry the Dirac propagator
but, instead, that due to Klein and Gordon. This endows the new field
with mass dimension one, rather than three half as for the Dirac field.
This happens despite the fact that the new field is fermionic,
and carries spin one half. It is endowed with several novel features. 
A few of
these are summarised in  Table I.

The first two properties enumerated in the second column are in accord to 
formal expectations based on Wigner's
work~[\refcite{Wigner:1964ep}], 
while da Rocha and Rodrigues
have shown~[\refcite{daRocha:2005ti}] that the new field is class 5 in 
Lounesto spinor classification~[\refcite{Lounesto:1997pl}]. 
The field carries a `minimal' element of non-locality
as required by the 1966 work of Lee and Wick~[\refcite{Lee:1966ik}] 
on Wigner classes and may be important for
resolving   the `horizon problem' in cosmology.

The third and the fourth results are intertwined. In contrast,
they have emerged as a surprise. Unexpectedly.
The mass dimensionality of the new field immediately places it
out of bounds for essentially all standard model interactions. 
The new particles cannot be part of the standard model 
fermionic doublets because the mass dimensionalities mismatch.
The most dominant interaction which the new particles
can carry is with the Higgs~[\refcite{f7}].

In addition, and unlike the standard model fermions where
the quartic self interaction is suppressed by two powers of Planck mass,
the new particles  carry an unsuppressed, and 
crucial~[\refcite{Ahn:2004xt}], 
quartic self interaction. This is because of their mass
dimensionality one.
These aspects  endow the new particles with a darkness with
respect to detectors made of the terrestrial 
standard model material and makes
them natural dark matter candidates. 
Due to its over abundance with respect to the standard model
matter in the astrophysical and cosmological environments, 
the  gravitational effects of dark matter remain as its primary signature.
It is there that dark matter primarily 
manifests its presence and its consequences.

Given the role of Higgs in its interactions with the new dark
matter candidate one may expect its appearance at Higgs factories.
By the same token, its effects in early universe, 
particularly in the context of matter-antimatter asymmetry, 
may be profound. This is due to the mismatch of the $CPT$ phases
at the interface with the standard-model matter.  
But here, our emphasis is on obtaining the darkness. The new
construct provides it in a natural fashion without extending
the spacetime symmetries, say to super-symmetry.

Lest these results acquire a mysterious life let us hasten to add
a few remarks to help evaporate any such feeling.
The entire physical content of the seventy six pages of
published calculations and discussions 
in 
references~[\refcite{Ahluwalia-Khalilova:2004sz,Ahluwalia-Khalilova:2004ab}] 
essentially resides in the fact that the  $C$, $P$, and $T$ properties,
and the emergent propagators, of the quantum fields
is carried by the  \textit{relative helicities and
phases} of the $(1/2,0)$- and $(0,1/2)$- transforming components of the
$(1/2,0)\oplus(0,1/2)$ spinors.  For the eigenspinors of the 
$P$ (i.e., spinors underlying the Dirac field), the relative helicities
of the said components are \textit{same}.
For the eigenspinors of the 
$C$ (i.e., the spinors underlying the new spin one half field), 
the relative helicities
of the said components are \textit{opposite}. The remaining degree of freedom
in the relative phases between these spinorial components is fixed
by requiring that these be such that they become $P$ and $C$ eigenspinors.
 Once this is appreciated the eigenspinors in the rest frame, apart from
a normalisation factor, are uniquely determined. The $(1/2,0)\oplus(0,1/2)$
boost then furnishes us a set of complete spinors. One set of four
spinors for the Dirac case, and the other for the new construct. The 
propagator then turns out to depend on the respective spin sums. For the 
Dirac case these spin sums have the usual $\frac{1}{m}\gamma^\mu p_\mu
\pm \openone$ form and yield
the celebrated Dirac propagator. 
For the new field these spin sums take
the form $\openone\pm \mathcal{G}(\mathbf{p})$, and yield the 
propagator 

\begin{equation}
\mathcal{S} (x-x^\prime) = \int\frac{d^4 p}{(2\pi)^4}\;
\mathrm{e}^{i p_\mu(x^\mu-x^{\prime\mu})}\;  
\frac{\openone+\mathcal{G}(\mathbf{p})}{p_\mu p^\mu - m^2 + i\epsilon}
\label{eq:noteworthy}
\end{equation}
The limit $\epsilon\to0^+$ is understood. 
A simple calculation using the eigenspinors
of $C$ shows that $\mathcal{G}(\mathbf{p})$
is an odd function of $\mathbf{p}$: 
$\mathcal{G}(\mathbf{p})= - \,\mathcal{G}(- \mathbf{p})$.
This has the consequence that in the absence of a preferred direction
\textemdash~ a circumstance which may be spoiled by existence, say,
of an external gravitomagnetic field \textemdash~the integration over
$\mathcal{G}(\mathbf{p})$ vanishes. This leaves behind a 
Klein-Gordon propagator multiplied by an identity matrix in the
$(1/2,0)\oplus(0,1/2)$ representation space. 

\begin{quote}
{\textbf{Table I.} Some of the properties of the new field
as compared to that of Dirac. Here, $\openone$ represents
a $4\times 4$ 
identity matrix in the $(1/2,0)\oplus(0,1/2)$ representation space.
For the commutator and anti-commutator relations it is implicit, as
a knowledgeable reader knows, that their validity holds while acting
on the relevant
spinors. Same holds for the statement on $(CPT)^2$. The 
propagators are obtained by calculating the vacuum expectation
value of the relevant time ordered product of the 
field operators. The calculational details 
appear 
in~[\refcite{Ahluwalia-Khalilova:2004sz,Ahluwalia-Khalilova:2004ab}].}
\end{quote}
\begin{table}
\begin{center}
\begin{tabular}{|l|l|}
\hline
\textit{Quantum field based on} & \textit{Quantum field based on}\\
\textit{Eigenspinors of P~[\refcite{Dirac:1928hu}]}      
& \textit{Eigenspinors of C~[\refcite{Ahluwalia-Khalilova:2004sz,Ahluwalia-Khalilova:2004ab}]}
     \\ \hline
$\{P,C\} =0 $          & $[P,C]=0$              \\
$(CPT)^2 = + \openone $        & $(CPT)^2 = - \openone $         \\
Propagator: Dirac      & Propagator: Klein-Gordon $\times \openone$  \\
Mass dimension: ${3}/{2}$    & Mass dimension: 1      \\ 
\hline
\end{tabular}\label{Table: I}
\end{center}
\end{table}

Should the reader 
wonder as to where did the Lagrangian enter; the answer is,
nowhere. And, that is the beauty of Wigner-like arguments we
constructed. It is now obtained by `inverting' the propagator.
Similarly, the new construct does not apply to Majorana neutrinos.
Was that to be done, they would cease to be part of the 
standard model doublets due to mass dimensionality carried by
the new construct. A more detailed discussion on the subject
can be found in the already cited references. Here it suffices to note
that the 1937 Majorana proposal is not at the 
level of representation space~[\refcite{Majorana:1937vz}]. 
It still uses Dirac spinors. Only
at the level of Fock space does
it identify the `particle-' and `antiparticle-'   
creation and destruction operators.

\vspace{9pt}

\noindent
\section{Conclusion} 
If one insists on the validity of the
general relativistic description of gravitation on astrophysical and
cosmological length scales, then the
considered  Friedmann-Robertson-Walker cosmology 
with Einsteinian cosmological constant, and vanishing spatial curvature
constant, 
unambiguously requires significant amount of dark matter. This requirement
is consistent with other indications for dark matter. 
The same spacetime symmetries that
underlie the freely falling frames of Einsteinian gravity also provide
symmetries, which for the spin one half representation
space, furnish a novel construct that carries extremely limited interactions
with respect to the terrestrial detectors made of the matter
described by the  standard model of high energy physics. 
A case is made that both the
`luminous' and the `dark' matter are  residents of the same representation
space but they derive their respective `luminosities' and `darkness' 
from being either the $(CPT)^2 = + \openone$ carrying 
eigenstates of the
parity operator,
or the  $(CPT)^2 = - \openone$ carrying 
eigenstates of the charge conjugation operator.

\vspace{9pt}

\section*{Acknowledgments} 
I thank Daniel Grumiller 
for 
our ongoing discussions and for his permission 
to include some of the  wisdom of our joint publications
on darkness of dark matter in this essay. I thank Garth Antony Barber 
for our discussion on  CosmoCoffee~[\refcite{Barber:2006ga}].
It helped me to make some of the claims more precise. 
It is also a pleasure to thank Thanu Padmanabhan 
for his 
constructive remarks on the first draft of this essay.
To the younger blood, which remains anonymous,
thank you for your fan mail.
Many  of these ideas crystallised, and took a new form, during
our February-March vacation in 
Australia. For that I thank Karan Ahluwalia, his family, and my wife, 
for creating that wonderful space and time where one
can think in an inspired manner.

Much of this work was done at the Department of Mathematics, 
University of Zacatecas (Mexico). Extended hospitality of that 
institution, and in particular that of its Director Gema Mercado,
is greatfully acknowledged.



\small
\begin{thebibliography}{99}

\bibitem{Persic:1995ru}
  M.~Persic, P.~Salucci and F.~Stel,
  \textit{The Universal rotation curve of spiral galaxies: 1. The dark matter
  connection,}
  Mon.\ Not.\ Roy.\ Astron.\ Soc.\  {\bf 281}, 27 (1996)
  [arXiv:astro-ph/9506004].

\bibitem{Wittman:2000tc}
  D.~M.~Wittman, J.~A.~Tyson, D.~Kirkman, I.~Dell'Antonio and G.~Bernstein,
  \textit{Detection of weak gravitational lensing distortions of distant galaxies by
  cosmic dark matter at large scales,}
  Nature {\bf 405}, 143 (2000)
  [arXiv:astro-ph/0003014].

\bibitem{Hoekstra:2003pn}
  H.~Hoekstra, H.~K.~C.~Yee and M.~D.~Gladders,
  \textit{Properties of galaxy dark matter halos from weak lensing,}
  Astrophys.\ J.\  {\bf 606}, 67 (2004)
  [arXiv:astro-ph/0306515].




\bibitem{Milgrom:1983pn}
  M.~Milgrom, 
    \textit{A modification of the {N}ewtonian dynamics: Implications for
   galaxies,}
  Astrophys.\ J.\ {\bf 270}, 371 (1983).

\bibitem{Sanders:2002pf}
  R.~H.~Sanders and S.~S.~McGaugh, 
     \textit{Modified {Newtonian} dynamics as an
   alternative to dark matter,}  
  Ann.\ Rev.\ Astron.\ Astrophys.\ {\bf 40}, 263 (2002)
  [astro-ph/0204521].


\bibitem{Bekenstein:2004ne}
  J.~D.~Bekenstein, 
    \textit{Relativistic gravitation theory for the {MOND}
    paradigm,}  {Phys. Rev.} {\bf D70},  083509 (2004) 
  [astro-ph/0403694]. J.
  D. Bekenstein, Phys.\ Rev.\ D.\ {\bf 71}, 069901 (2005) (erratum).

\bibitem{Zhao:2005gk}
  H.~S.~Zhao and B.~Famaey,
  \textit{Refining MOND interpolating function and TeVeS Lagrangian,}
  Astrophys.\ J.\  {\bf 638} (2006) L9
  [arXiv:astro-ph/0512425].

\bibitem{Sobouti:2006rd}
  Y.~Sobouti,
  \textit{An $f(R)$ gravitation instead of dark matter,}
  arXiv:astro-ph/0603302.


\bibitem{Ahluwalia-Khalilova:2006ez}
  D.~V.~Ahluwalia-Khalilova,
  \textit{Dark matter: A phenomenological existence proof,}
  arXiv:astro-ph/0601489.

\bibitem{Vishwakarma:2006dj}
  R.~G.~Vishwakarma,
  ``Comments on Dark matter: A phenomenological existence proof,''
  arXiv:astro-ph/0603213.

\bibitem{Ahluwalia-Khalilova:2006ky}
  D.~V.~Ahluwalia-Khalilova,
  ``A reply to a comment on 'Dark matter: A phenomenological existence
  proof',''
  arXiv:astro-ph/0603256.


\bibitem{Ahluwalia-Khalilova:2004sz}
  D.~V.~Ahluwalia-Khalilova and D.~Grumiller,
  \textit{Dark matter: A spin one half fermion field with mass dimension one?,}
  Phys.\ Rev.\ D {\bf 72} (2005) 067701
  [arXiv:hep-th/0410192].

\bibitem{Ahluwalia-Khalilova:2004ab}
  D.~V.~Ahluwalia-Khalilova and D.~Grumiller,
  \textit{Spin half fermions with mass dimension one: 
  Theory, phenomenology, and dark matter,}
  JCAP {\bf 0507} (2005) 012
  [arXiv:hep-th/0412080].

\bibitem{Perlmutter:1997zf}
  S.~Perlmutter {\it et al.}  [Supernova Cosmology Project Collaboration],
  \textit{Discovery of a supernova explosion at half the age of the universe 
	  and its cosmological implications,}
  Nature {\bf 391} (1998) 51
  [arXiv:astro-ph/9712212].


\bibitem{Riess:1998cb}
  A.~G.~Riess {\it et al.}  [Supernova Search Team Collaboration],
  \textit{Observational evidence from supernovae for an accelerating 
   universe and a
  cosmological Constant,}
  Astron.\ J.\  {\bf 116}, 1009 (1998)
  [arXiv:astro-ph/9805201].

\bibitem{Perlmutter:1998np}
  S.~Perlmutter {\it et al.}  [Supernova Cosmology Project Collaboration],
  \textit{Measurements of $\Omega$ and $\Lambda$ from 42 
	high-redshift supernovae,}
  Astrophys.\ J.\  {\bf 517}, 565 (1999)
  [arXiv:astro-ph/9812133].

\bibitem{Tonry:2003zg}
  J.~L.~Tonry {\it et al.}  [Supernova Search Team Collaboration],
  \textit{Cosmological results from high-z supernovae,}
  Astrophys.\ J.\  {\bf 594} (2003) 1
  [arXiv:astro-ph/0305008].

\bibitem{Riess:2004nr}
  A.~G.~Riess {\it et al.}  [Supernova Search Team Collaboration],
  \textit{Type Ia Supernova discoveries at $z>1$ from the Hubble Space 
  Telescope:
   Evidence for past deceleration and constraints 
    on dark energy evolution,}
  Astrophys.\ J.\  {\bf 607}, 665 (2004)
  [arXiv:astro-ph/0402512].

\bibitem{Astier:2005qq}
  P.~Astier {\it et al.},
  \textit{The Supernova Legacy Survey: Measurement 
   of $\Omega_m$, $\Omega_\Lambda$ and $w$
   from the first year data set,}
  arXiv:astro-ph/0510447. 


\bibitem{Jassal:2005qc}
  H.~K.~Jassal, J.~S.~Bagla and T.~Padmanabhan,
  \textit{Observational constraints on low redshift evolution of dark energy: How
  consistent are different observations?,}
  Phys.\ Rev.\ D {\bf 72} (2005) 103503
  [arXiv:astro-ph/0506748].


\bibitem{Jassal:2006gf}
  H.~K.~Jassal, J.~S.~Bagla and T.~Padmanabhan,
  \textit{The vanishing phantom menace,}
  arXiv:astro-ph/0601389.








\bibitem{Spergel:2003cb}
  D.~N.~Spergel {\it et al.}  [WMAP Collaboration],
  \textit{First year Wilkinson Microwave Anisotropy Probe (WMAP) observations:
  Determination of cosmological parameters,}
  Astrophys.\ J.\ Suppl.\  {\bf 148} (2003) 175
  [arXiv:astro-ph/0302209].




\bibitem{Copeland:2006wr}
  E.~J.~Copeland, M.~Sami and S.~Tsujikawa,
  \textit{Dynamics of dark energy,}
  arXiv:hep-th/0603057.

\bibitem{Wiltshire:2005fw}
  D.~L.~Wiltshire,
  \textit{Viable inhomogeneous model universe without dark energy 
         from primordial inflation,}
  arXiv:gr-qc/0503099.

\bibitem{Alnes:2006uk}
  H.~Alnes and M.~Amarzguioui,
  The supernova Hubble diagram for off-center observers in a spherically
  symmetric inhomogeneous universe,''
  arXiv:astro-ph/0610331.

\bibitem{Mustapha:1997xb}
  N.~Mustapha, B.~A.~Bassett, C.~Hellaby and G.~F.~R.~Ellis,
  ``Shrinking II -- The Distortion of the Area Distance-Redshift Relation in
  Inhomogeneous Isotropic Universes,''
  Class.\ Quant.\ Grav.\  {\bf 15} (1998) 2363
  [arXiv:gr-qc/9708043].


\bibitem{Rasanen:2006kp}
  S.~Rasanen,
  ``Accelerated expansion from structure formation,''
  JCAP {\bf 0611} (2006) 003
  [arXiv:astro-ph/0607626].

\bibitem{Carter:2005mu}
  B.~M.~N.~Carter, B.~M.~Leith, S.~C.~C.~Ng, A.~B.~Nielsen and D.~L.~Wiltshire,
  \textit{Exact model universe fits type IA supernovae data with no cosmic
  aacceleration,}
  arXiv:astro-ph/0504192.



\bibitem{f2}For additional discussion,  
see, e.g.,~\refcite{Alnes:2005rw,Buchert:2005kj}. 

\bibitem{Alnes:2005rw}
  H.~Alnes, M.~Amarzguioui and O.~Gron,
  \textit{An inhomogeneous alternative to dark energy?,}
  arXiv:astro-ph/0512006.

\bibitem{Buchert:2005kj}
  T.~Buchert,
  \textit{On globally static and stationary cosmologies with or without a
   cosmological constant and the dark energy problem,}
  Class.\ Quant.\ Grav.\  {\bf 23} (2006) 817
  [arXiv:gr-qc/0509124].




\bibitem{Gron:2002ox}
  O.~Gron, 
   \textit{A new standard model of the universe,}
    Eur.\ J. Phys. {\bf 23} (2002) 135-144.


\bibitem{Frieman:1994gf}
  J.~A.~Frieman,
  \textit{The standard cosmology,}
  arXiv:astro-ph/9404040.



\bibitem{Lake:2006fb}
  K.~Lake,
  \textit{Integration of the Friedmann equation for universes of arbitrary
  complexity,}
  arXiv:gr-qc/0603028.

\bibitem{f3}
If $w^\Lambda = {p^\Lambda}/{\rho^\Lambda} $
is confined to  the choice
$w^\Lambda=-1$ then a time-independent  $\rho^\Lambda$ 
corresponds to the Einsteinian  cosmological constant.

\bibitem{f4}
This result is
under the assumption that one may discard negligible contribution coming
from radiation and from the radiation-dominated epoch.




\bibitem{Kolb:1990Turner}
E.~W.~Kolb and M~S.~Turner, \textit{The early universe} (Westview Press, 1990) 


\bibitem{Freedman:2000cf}
  W.~L.~Freedman {\it et al.},
  \textit{Final results from the 
   Hubble Space Telescope Key Project to measure the
  Hubble constant,}
  Astrophys.\ J.\  {\bf 553} (2001) 47
  [arXiv:astro-ph/0012376].

\bibitem{Chaboyer84:2001}
B.~Chaboyer,
  \textit{Astrophysical ages and time scales,} in: T. von Hippel,
C. Simpson, N. Manset (Eds.),
APS Conference Series, Vol. 245, 2001, p.162.

\bibitem{f5}
In a recent
preprint Balasin and Grumiller
have shown~\refcite{Balasin:2006cg} 
that Newtonian arguments over-estimate 
the amount of matter needed to explain non-Keplerian 
galactic rotation curves by more than thirty percent. This 
decrease in demand on dark matter content would need to be accounted for
in more careful estimation of dark matter from the galactic rotation
curves and gravitational lensing data. For care that needs
to be exercised in such joint analysis, see
 Faber-Visser preprint~\refcite{Faber:2005xc}. 

\bibitem{Balasin:2006cg}
  H.~Balasin and D.~Grumiller,
  \textit{Significant reduction of 
	galactic dark matter by general relativity,}
  arXiv:astro-ph/0602519.


\bibitem{Faber:2005xc}
  T.~Faber and M.~Visser,
  \textit{Combining rotation curves 
   and gravitational lensing: How to measure the
  equation of state of dark matter in the galactic fluid,}
  arXiv:astro-ph/0512213.



\bibitem{Weinberg:1995sw}
S.~Weinberg, \textit{The quantum theory of fields, Vol. I} (Cambridge 
University Press, Cambridge, 1995).

\bibitem{Ryder:1996lh}
L.~H.~Ryder, \textit{Quantum field theory}
(Cambridge University Press, Cambridge, 1997).

\bibitem{Wigner:1964ep}
E.~P.~Wigner, in \textit{Group theoretical concepts and methods
in elementary particles physics: Lectures of the Istanbul summer school
of theoretical physics (July 16-August 4, 1962),} pp. 37-80, 
Editor F.~G\"ursey
(Gordon and Breach, New York, 1964). 

\bibitem{f6}
For work on spin one, see Ref.~\refcite{Ahluwalia:1993zt}.



\bibitem{Ahluwalia:1993zt}
  D.~V.~Ahluwalia, M.~B.~Johnson and T.~Goldman,
  \textit{A Bargmann-Wightman-Wigner type quantum field theory,}
  Phys.\ Lett.\ B {\bf 316} (1993) 102
  [arXiv:hep-ph/9304243].

\bibitem{Dirac:1928hu}
  P.~A.~M.~Dirac,
  \textit{The quantum theory of electron,}
  Proc.\ Roy.\ Soc.\ Lond.\ A {\bf 117} (1928) 610.




\bibitem{daRocha:2005ti}
  R.~da Rocha and W.~A.~J.~Rodrigues,
  \textit{Where are ELKO spinor fields in 
   Lounesto spinor field classification?}'
  Mod.\ Phys.\ Lett.\ A {\bf 21} (2006) 65
  [arXiv:math-ph/0506075].

\bibitem{Lounesto:1997pl}
P.~Lounesto, \textit{Clifford algebras and spinors}
(Cambridge University Press, Cambridge, 1997).




\bibitem{Lee:1966ik}
  T.~D.~Lee and G.~C.~Wick,
  \textit{Space inversion, time reversal, 
   and other discrete symmetries in local
  field theories,}
  Phys.\ Rev.\  {\bf 148} (1966) 1385.

\bibitem{f7}
See, references~\refcite{Ahluwalia-Khalilova:2004sz,Ahluwalia-Khalilova:2004ab} 
for additional discussion.

\bibitem{Ahn:2004xt}
  K.~J.~Ahn and P.~R.~Shapiro,
  \textit{Formation and evolution of the self-interacting dark matter 
   halos,}
  Mon.\ Not.\ Roy.\ Astron.\ Soc.\  {\bf 363} (2005) 1092
  [arXiv:astro-ph/0412169].


\bibitem{Majorana:1937vz}
  E.~Majorana,
  \textit{Theory of the symmetry of electrons and positrons,}
  Nuovo Cim.\  {\bf 14} (1937) 171. An English translation
  of this 1937 classic Italian language paper   by  L. Maiani
  can be found in  Soryushiron\ Kenkyu {\bf 63} (1981) 149.
  The reader should feel free to request a pdf file from the
  author of this essay.


\bibitem{Barber:2006ga} 
G.~Barber~and~D.~V.~Ahluwalia-Khalilova,\\
http://cosmocoffee.info/viewtopic.php?p=1409\#1409.




\end{thebibliography}
\end{document}